\documentstyle[11pt,aaspp4]{article}

\begin{document}

\title{Reactivation and Precise IPN Localization of the Soft Gamma Repeater SGR1900+14}

\author{K. Hurley}
\affil{University of California, Berkeley, Space Sciences Laboratory,
Berkeley, CA 94720-7450}
\authoremail{khurley@sunspot.ssl.berkeley.edu}
\author{C. Kouveliotou} 
\affil{Universities Space Research Association at NASA Marshall Space Flight Center, 
ES-84, Huntsville AL 35812}
\author{P. Woods\altaffilmark{1}}
\affil{NASA Marshall Space Flight Center, ES--84, Huntsville AL 35812}
\author{T. Cline, P. Butterworth}
\affil{NASA Goddard Space Flight Center, Code 661, Greenbelt, MD 20771}
\author{E. Mazets, S. Golenetskii, D. Frederics}
\affil{Ioffe Physical-Technical Institute, St. Petersburg, 194021, Russia}

\altaffiltext{1}{Dept. of Physics, University of Alabama in Huntsville,
Huntsville, AL 35899}

\begin{abstract}

In 1998 May, the soft gamma repeater SGR1900+14 emerged from several years
of quiescence and emit a series of intense bursts, one with a time history
unlike any previously observed from this source.  Triangulation using \it Ulysses \rm, 
BATSE, and KONUS data give a 1.6 square arcminute error box near the
galactic supernova remnant G42.8+0.6.  This error box contains a
quiescent soft X-ray source which is probably a neutron star associated with the soft
repeater.

\end{abstract}

\keywords{gamma rays: bursts; stars: neutron; supernova remnants}

\section{Introduction}

The soft gamma repeaters (SGRs) differ from the classical gamma-ray bursts by their
durations (typically 0.1--1 s), soft spectra ($\rm kT \sim 35 \, keV$), and their
repetition.  Four SGRs are now known, and there is evidence that all of them are associated
with supernova remnants (SNRs), and are therefore neutron stars.  The source of the
well-known 1979 March 5 burst, SGR0525-66, is consistent with the position of
the N49 supernova remnant in the Large Magallanic Cloud  
(Cline et al. 1982).  SGR1806-20 (Atteia et al. 1987) is associated with the  
SNR G10.0-0.3, located towards the galactic center (Kulkarni \& 
Frail 1993; Kouveliotou et al. 1994; Kulkarni et al. 
1994; Murakami et al. 1994).  SGR1627-41 may be associated with
G337.0-0.1 (Hurley et al. 1998a,b; Woods et al., 1998). 

SGR1900+14 is located near the galactic plane and 
was, until 1998 May, the least active of the SGRs (Hurley 1996), hindering
attempts to locate the source accurately.  Two possible 
small error boxes for SGR1900+14 were obtained by Hurley et al. (1994) 
with the network synthesis method, using the Interplanetary Network (IPN) data 
from \it Ulysses \rm and BATSE.  The location of the first is very close to the 
SNR G42.8+0.6, suggested as one of two possible SNRs associated with this object
by Kouveliotou et al. (1994). Both the ROSAT sky survey (Vasisht et al. 1994) and a pointed observation 
at this position (Hurley et al. 1996a) indicated that a point-like quiescent X-ray source was associated 
with this error box.  Optical and infrared observations of the X-ray position revealed a 
peculiar double M star system (Vrba et al. 1996).  No ROSAT source was detected in the second error box, which lies 3.75$\rm^o$ from the first.   
(Li et al. 1997).  SGR1900+14 underwent
an extraordinary resumption of activity starting in 1998 May and triangulation
using IPN data (\it Ulysses \rm, BATSE, and KONUS-Wind in this case) has now reduced the possible source location to a single, 1.6 square arcminute error
box consistent with the ROSAT quiescent source.

\section{Source Activity}

SGR1900+14 was discovered when it burst three times in 1979 (Mazets et al. 1979).  It
was not heard from again until 1992, when it emitted four bursts (Kouveliotou et al. 1993).
It then lay dormant again until 1998 May 26, when it began a series of more than 50 bursts over
the next several months (Kouveliotou et al. 1998a), culminating in the giant flare of 1998 August 27 (Hurley et al. 1998c).  The source remains active as of this writing.  In addition to these events, which were confirmed either by their BATSE and/or IPN localizations, over 30
bursts have occurred which are suspected to originate
from this source, but for which no locations can be derived.  It is also possible
that numerous weak bursts could have occurred during periods of BATSE Earth occultation,
which would have been below the thresholds of \it Ulysses \rm and KONUS-Wind.

The time histories of most of these events consisted of a single peak with
duration $\lesssim$ 100 ms (figure 1).  However, on 1998 May 30 the source emitted a
320 s long burst consisting of multiple peaks, many with structures and/or durations
longer than 100 ms (Kouveliotou et al. 1998a; figure 1).  "Bunching" of bursts had
previously been reported from SGR1806-20 (Kouveliotou et al. 1996).  However, whereas
the bursts from SGR1806-20 were observed by the Rossi X-Ray Timing Explorer PCA in the
2-60 keV range, and came at a rate of $\lesssim$1 per minute, the bunching observed
from SGR1900+14 was in the hard X-ray range, 25-300 keV, and came at a rate up to
one every several seconds.  At least three BATSE events during this period reached peak
25 - 300 keV luminosities of $\sim 1000 L_{Edd}$, considerably greater than the
XTE events.  Thus this type of event appears to be unique to SGRs.

\section{Source location}

We have localized this source by triangulation.  Specifically, each Ulysses event for which
there was a BATSE and/or a KONUS-Wind response was used to produce an annulus of possible
arrival directions.  The method is explained in more detail in Hurley et al. (1998d),
and Table 1 lists the events triangulated to date.    
However, there are two aspects which make the localization of SGRs in general,
and this SGR in particular, unique.  First,
it is the first time since the launch of \it Ulysses \rm in 1990 that the bursts from
SGR1900+14 have been intense
enough to trigger the GRB detector on this spacecraft, providing 32 ms time resolution data
which can be used for precise triangulation, and second, this
period of activity lasted long enough that the changing direction of the \it Ulysses \rm - Earth vector, which is the center of the triangulation annulus, 
made it possible to derive annuli which crossed one another, generating a small error box.
For simplicity, we have defined this error box with only two $\sim 24 \arcsec$ wide annuli, one
from 1998 May 26, and the other from 1998 July 19.  Although all the
events in Table 1 produced annuli which are consistent with it, they are in some cases
redundant, and in others, wider than the two we have selected due to low time resolution
data.

Figure 2 shows the new localization and the VLA \footnote{
The Very Large Array is a facility of the National Science Foundation
operated under cooperative agreement by Associated Universities, Inc.} radio contours of G42.8+0.6.  The ROSAT quiescent X-ray source error circle is also shown.  This
error box is consistent with the previous ones,
but the area has decreased from $\approx$3.5 square degrees (Mazets et al. error box quoted
in Kouveliotou et al. 1993) to $\approx$10 square arcminutes (for the two widely
separated network synthesis error boxes),
to $\approx$1.6 square arcminutes. (for the single error box derived from the recent source activity).  Table 2
gives the coordinates of the conservatively estimated 3 $\sigma$
confidence error box corners.  The error box center is at $\alpha(2000)=286.8120, \delta(2000)=9.3283^o$.

It has been suggested that the superluminal galactic source GRS1915+105 is responsible for the bursts from SGR1900+14 (Mirabel 1994a,b).  The error box in Table 2, however, lies
$\rm \sim 2.5^o $ from this source, definitively excluding it as a candidate.

\section{Discussion}

Since an SNR-SGR connection exists for three SGRs,
it seems likely that SGR1900+14 is also associated with an SNR.  Similarly,
since at least two SGRs have quiescent soft X-ray sources associated with them
(SGR0525-66 and SGR1806-20), it is
likely that the ROSAT source in the error box of SGR1900+14 is associated with
this SGR (further evidence for this association is presented in Hurley et al.
1998e).  The X-ray source is clearly not located within
the radio contours of G42.8+0.6 (figure 2), and this poses an interesting
question.  If the two are associated, the neutron star would have to have
a transverse velocity
v(km/s)=276 D(kpc) $\theta$(arcminutes)/A (kyr)
where D is the distance to the supernova remnant, $\theta$ is the angular
separation between the site of the supernova explosion and the neutron star,
and A is the remnant's age (e.g. Hurley et al. 1996b).  The actual velocity
might be $\approx \sqrt{3/2}$ this much.  The distance to G42.8+0.6 has
been estimated as 5 kpc using the luminosity-diameter relation (e.g Vasisht et 
al. 1994), which is
probably uncertain by a factor of 2.  However, it agrees with an independent
estimate from the n$_H$ as measured from the quiescent X-ray spectrum (Hurley
et al. 1998e).  The angular separation is difficult to
judge because G42.8+0.6 is highly asymmetrical.  Reasonable estimates might
range from 7 - 20 $\arcmin$.  Finally, the age is probably less than 20,000 y,
or the SNR would not be detectable (Braun, Goss, and Lyne 1983), and might
be as young as 5,000 y.  This leads to a wide range of possible velocities,
from 480 to 5,500 km/s, even the lowest of which are rather high.  This may
be compared to the estimated velocity of SGR0525-66, which is 1200 km/s (Thompson
and Duncan 1995), and that of SGR1806-20, $>$500 km/s (Kulkarni et al. 1994).
How did the neutron star acquire this velocity?

Elsewhere, evidence has been presented that SGR1900+14 may be a magnetar (Kouveliotou 
et al. 1998c; Hurley et al., 1998c), i.e. a neutron star with a field strength
in excess of 10$^{14}$ G.  If the magnetar model is basically correct, it may
provide an explanation for the high velocity of this object (Thompson and Duncan 1995):
the strong field may bring about an anisotropy in the neutrino radiation of the
newly formed neutron star.  Moreover, since magnetar lifetimes are expected to
be of the order of 10,000 years, they are detectable over shorter times than
supernova remnants.  Thus the magnetar model constrains one to accept the
hypothesis that SGR1900+14 originated in the SNR G42.8+0.6.  The only other
possibility is that it originated in an SNR which is presently undetectable,
contradicting the lifetime arguments.

As SGR1900+14 remains active, it is possible that this error box can be refined
further in the future.  Such error boxes will provide interesting calibrations
of the triangulation technique.  However, in view of the present error box, as well
as the 5.16 s periodicity observed in the quiescent X-ray source (Hurley et al. 1998e)
and the bursting SGR (Hurley et al. 1998c), the association between the quiescent
source and the SGR may be considered to be secure.

\acknowledgments
KH is grateful to NASA for support under the CGRO 
Guest Investigator Program (NAG 5-1560), and to JPL for support of Ulysses (Contract 
958056).  CK acknowledges
support under the CGRO guest investigator program (NAG 5-2560).  We are indebted
to D. Frail for providing the VLA map in figure 2.
At the Ioffe Institute this work was partially supported by an RSA contract and by
RFBR grant 96-02-16860.

\newpage

\begin{figure}
\plotone{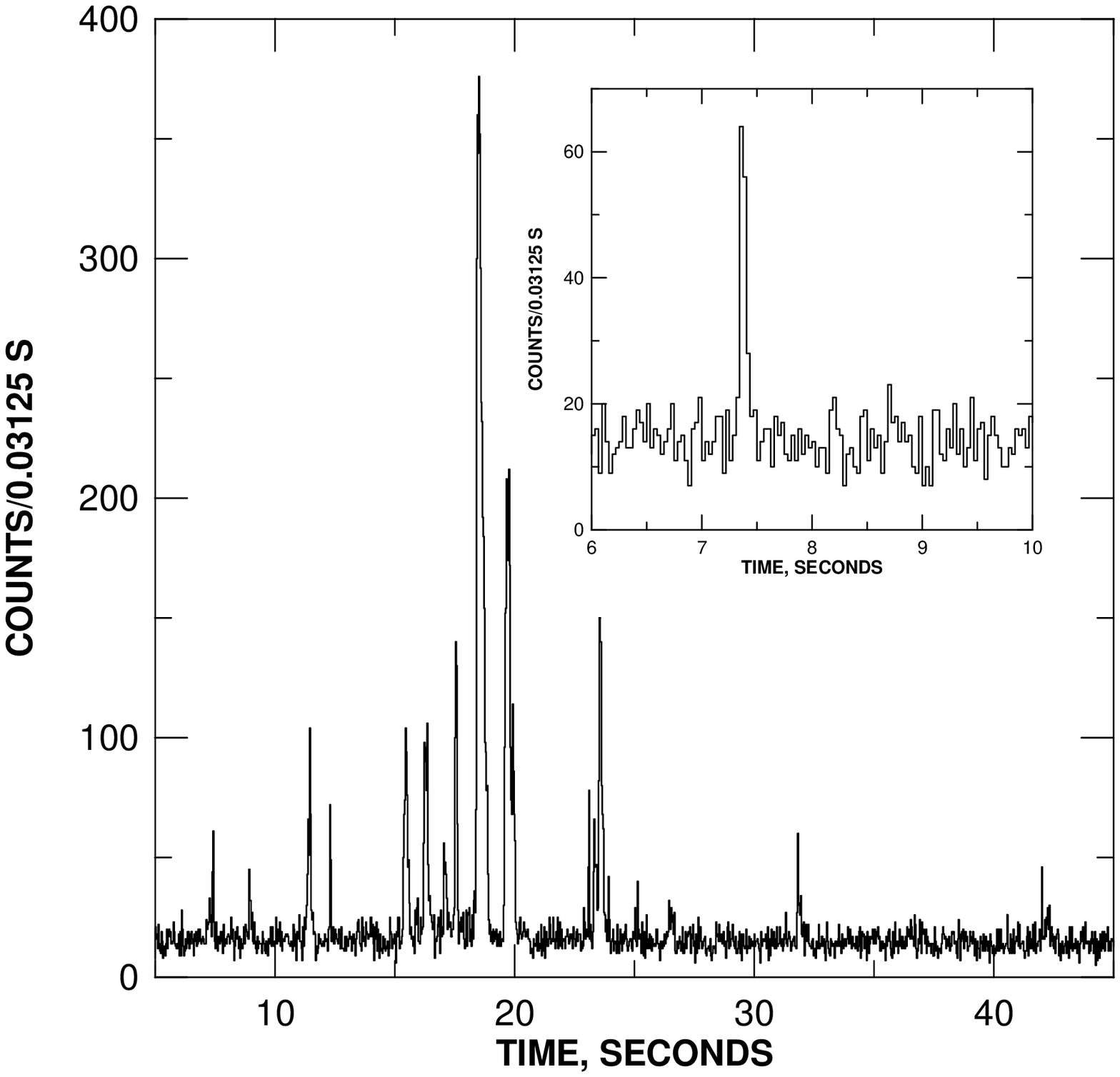}
\caption{\it Ulysses \rm 25-150 keV time history of a portion of the unusual event of
1998 May 30.  This type of time history has never been observed from 
SGR1900+14 before.  
Inset:  \it Ulysses \rm 25-150 keV time history of a single-spike burst from
SGR1900+14 on 1998 May 26.  This time history is typical of this source and
most SGR bursts. \label{fig1}
}
\end{figure}

\begin{figure}
\plotone{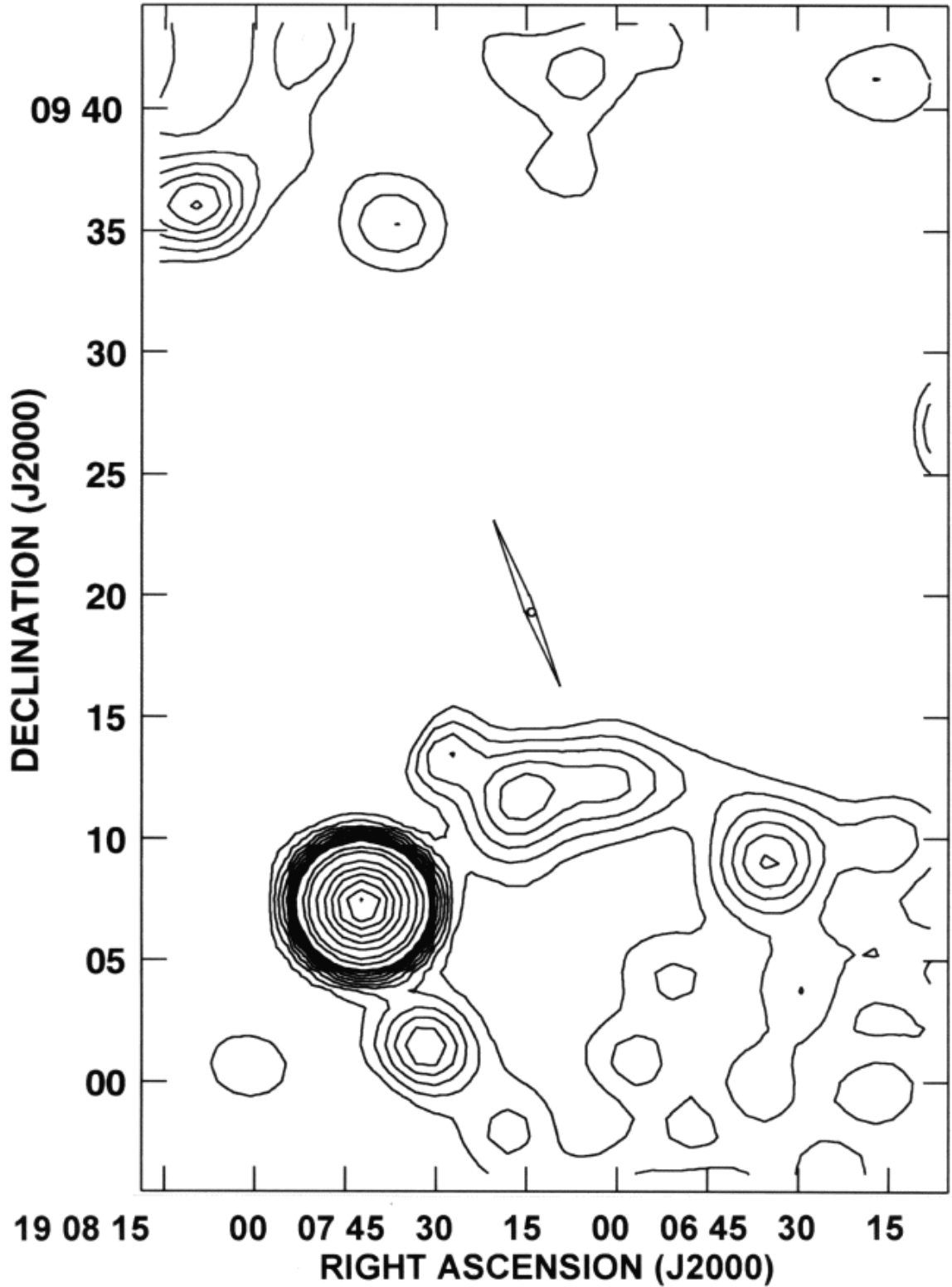}
\caption{The error box for SGR1900+14 superimposed on a radio map of this
region.  The radio continuum image was taken at the VLA on 1994 February 07
at a frequency of 327 MHz (Vasisht et al. 1994).  The synthesized beamsize is approximately
3 arcminutes.  Contour levels
progress in steps of 20 mJy/beam up to 200 mJy/beam.  Higher contour levels
progress in steps of 20 mJy/beam, to the peak of 1.6 Jy.  The position
of the ROSAT source within the error box is indicated.  \label{fig2}
}
\end{figure}

\clearpage

\begin{deluxetable}{cccccccccc}
\tablecaption{\it SGR1900+14 Bursts Triangulated by the IPN}
\tablehead{
\colhead{Date}&\colhead{Seconds, UT}&\colhead{Ulysses}&\colhead{BATSE}&
\colhead{KONUS-WIND}
}
\startdata
26 MAY 98 & 77429 & YES\tablenotemark{a} & RI\tablenotemark{b}  \nl
26 MAY 98 & 80646 & YES & RI   & YES \nl
27 MAY 98 & 15721 & YES & NO\tablenotemark{c}   & YES \nl
30 MAY 98 & 32624 & YES & YES  & YES \nl
30 MAY 98 & 34076 & RI  & RI   &    \nl
30 MAY 98 & 41332 & RI  & RI   & YES \nl
30 MAY 98 & 84457 & YES &     & YES \nl
 7 JUN 98 & 30749 & RI  & YES  & YES \nl
19 JUL 98 & 30383 & YES & YES  & YES \nl
19 JUL 98 & 60010 & YES & YES  & YES \nl
\enddata
\tablenotetext{a}{YES: Burst was observed in triggered, high time resolution mode}
\tablenotetext{b}{RI: Burst was observed as a rate increase, in low time resolution mode}
\tablenotetext{c}{NO: Data were available, but event was not observed}
\end{deluxetable}
\clearpage

\clearpage
\begin{table*}
\begin{center}
\begin{tabular}{cc}
$\alpha(2000), degrees$ & $\delta(2000), degrees$  \\
\tableline
   286.7895 &     9.2717 \\
   286.8135 &     9.3224  \\
   286.8108 &     9.3338  \\
   286.8349 &     9.3844 \\

\end{tabular}
\end{center}
\tablenum{2}
\caption{Error box corners for SGR1900+14}
\end{table*}

\end{document}